# Excel Modelling - Transparency, Auditing and Business Use

Susan Allan, Risk Analytics, Lloyds Banking Group

susan_allan@bankofscotland.co.uk

**ABSTRACT**

*Within Lloyds Banking Group the heritage HBOS Corporate division deals with Corporate loans, and is required to assess these loans for risk in accordance with the Basle Accord regulations. Statistical Risk Rating models are developed by the risk analysts to assess the obligors credit worthiness. It is necessary then to provide the bankers who originated the loan ('Relationship Managers' or RMs) with an assessment tool to generate the loan rating upon which they base their lending decisions. Heritage HBoS Corporate required a new model build system for holding its Risk Rating models in 2006 as a result of more complex models being created to comply with the Basle Accord.. The use of Excel was promoted by the IT department for a number of reasons; the Excel solution now in use is reviewed in this paper.*

1. **INTRODUCTION**

This paper discusses the benefits and considerations employed in the use of the new Excel-based model build system, looking first at the issues with visibility to users, then at the good-practice guides that have been used within the company.

The paper then looks at issues that have arisen through the development of the new system, and how processes have been reviewed as a result of its use.

2. **BACKGROUND**

When the RM has input the deal information in the front-end system ('Nexus') and generated a rating, this information then feeds into the models in the new model build system ('Nexus Model Designer' or NMD) to produce a rating for that loan, which feeds back into Nexus to be viewed by the RM. The loan is then passed via Nexus to the Sanctioners for review and approval, before the customer is advised of the loan price. The RMs and Sanctioners expressed a need to view the Risk Rating model when specific deals are entered so that they can query how a deal is being rated in a timely manner thus allowing a quick response to the customer.

3. **TRANSPARENCY**

In previous model build systems there had been no visibility of the risk models to anyone but the analysts permitted access to them. This meant that as models were developed there was no way of showing the RMs what they looked like without providing them with access to the system and subsequent training in it; given that the locations of some of these managers was spread around the country this was not practical.

It had become clear to the Risk Analytics model implementation team that there were a number of gaps in the current processes; as well as the ability to extend the visibility of models to the RMs throughout the model development, a full model audit trail was required by the internal and external auditors.

To this end the NMD Excel-based tool seemed to fit the requirements of being accessible to all, and using an existing system (Excel) which would require minimal training; the view was that most users had basic to moderate understanding of Excel functions.

**4. NMD: GOOD-PRACTICE GUIDELINES EMPLOYED**

Given the time already spent by the RMs in their 'day jobs' it was important that the users could work out in a short time period how their deals flow through the model and what the model is doing with each calculation. To this end good practice [Grossman & Ozluk 2003][Cernauskas, Kumiega, Van Vliet, 2007] [Bishop, 2006] meant using the following steps:

*4.1 Use a structured spreadsheet*

Best practice within HBoS follows the guidelines of having a structure which uses separate input, output and calculation sheets. This enables identification of changes to inputs/outputs and calculations, an important point for the system development to meet auditing requirements which are discussed in section 4.3. Within the NMD there is a 'formula walker' which enables the user to step through the formula from start to finish, and records the steps taken so that these can be passed to RMs/sanctioners to demonstrate how a deal gets its rating in a particular model.

The formula walker, shown below in Figure 1, has significantly improved the existing "Trace Precedents and Dependants" functionality by showing the sheet and cell name, formulae and values in one place.

Figure 1 – the NMD Formula Walker

*4.2 Ensure the formulae use common easily-interpreted functions*.

Where complex formulae exist it may be worth considering a 'wizard' to better detail what the formula is doing.

For example, in the HBOS Corporate models complex array formulae are used to conditionally map values across spreadsheets, and these are often difficult to interpret; more so if the conditional formula is extended with several conditions. To aid with this, a wizard was created in the NMD which builds such formula through a graphical interface; an example is shown in Figures 2a and 2b.

Figure 2a – Display of an array formula in the NMD

> =MAX (SecDI.ExposureResidualMaturity
> [SecDI.SecurityID = SEC_GteeADJ.SecurityID AND
> SecDI1.LinkFlag = 1 ] )
>
> The condition is shown here in the square brackets.  This says that this cell will take the maximum of the exposure Residual Maturity value from the column with these values in the SecDI sheet, if the security IDs for this security on this sheet (the SecGteeADJ sheet) and row match the security ID on the same row in the SecDI sheet, and also where the link flag on the SecDI sheet = 1 on the same row within the LinkFlag column.

Figure 2b – Standard Excel Array formula

> =MAX(IF(SecDI!$B$5:$B$754=SEC_GTEEADJ!$B5,
> IF(SecDI1!$C$5:$C$754=1,SecDI!$L$5:$L$754)))
>
> Same thing being said here but as an array formula – no names for the ranges, nor for the cells so interpretation is more difficult.  Naming on the columns/cells in the box above is made possible using Excel's named range function.

The wizard automatically creates both the user-friendly format (which is easier to interpret and explain to RMs) and the actual excel array formula on the spreadsheet.

### *4.3 Auditability*

In the NMD each time a change is made to the model it is logged.  This ensures that the internal and external audit requirement around transparency of change is covered.  To ensure that the regulatory requirements [Financial Services Authority, 2006] are met around the use of the rating system throughout the company and in the decisioning process, input is required from a number of parties (such as RMs, IT teams, and data teams who may need outputs named in a structured manner for the databases).

The NMD was built to log two types of changes – one where the internal model formula/lookups change, and the second where the inputs or outputs change.  For the first, a new 'revision' is logged when the amended model is saved.  For the second, a new version number is created when it is saved.  This new version number is referenced by the IT department when the model is promoted into the live environment for use with live deals.

With each change logged, the user name would be logged along with the time and date of the change in the NMD (Fig 3a).  It is then visible when the change was made, what the change entailed and who had made the change.

This is a step forward from the lack of transparency in previous systems, and is an important step in terms of ensuring the analysts can trace back to see why particular changes were entered.  With each change a description can be entered.  Given the number of iterations of models, versions can also be archived where appropriate within the NMD for recall at any time.

Fig. 3a – Version and revisions in NMD

Fig. 3b – Revision details

5. USER FEEDBACK AND LESSONS LEARNT

When the models are exported to the RMs, the assumption was that there would be sufficient knowledge of Excel to enable interrogation of the models as they are developed. Whilst functions such as the conditional mapping formulae can be explained, and were easily picked up by the RMs, there was some surprising feedback around the look and feel of the Excel view presented by the NMD models.

Fig. 4 – Visual Basic version showing model input screen

| Clear Inputs | Clear Qualitative Questions |
|---|---|
| **Customer Name:** | ABC |

| **Income Statement** | | (annual figures £000s) |
|---|---|---|
| Factor1 | (enter as +ve) | 11.000 |
| Factor2 | (enter actual value) | 124.000 |

| **Balance Sheet** | | (annual figures £000s) |
|---|---|---|
| Factor 3 | (enter as +ve) | 123.000 |
| Factor4 | (enter as +ve) | 123.000 |
| Factor5 | (enter as +ve) | 123.000 |
| Factor6 | (enter as +ve) | 123.000 |
| Factor7 | (enter as +ve) | 12.000 |

| **Other** | |
|---|---|
| Factor8 | 5 |
| Factor9 | FALSE |
| Factor10 | Average |
| Factor11 | 65 |
| Factor12 | 10101 |

**Calculations:**

SIC Code Model Category:     SIC code included in model build

Final Rating:     7

In previous simple models which had been replicated from the old model build system into Excel spreadsheets, much use had been made of Visual Basic (VB) coding to make them look glossy and easy to use for the RMs as shown in figure 4 above. The expectation from the RMs was that models from the NMD would have the same look as the simpler VB models they had seen previously. These required minimal interpretation and the expectation was that as both were Excel models there should be little difference. However with the NMD models, VB cannot be used as it is deemed by the IT designers as a corruptible program which cannot be controlled. Furthermore Microsoft will not support VB going forwards as there is a move to use **.**NET instead. As a result the look of models exported to the RMs from the NMD is simply that of a basic Excel spreadsheet with minimal formatting as depicted in Figure 5.

Figure 5 – NMD model input sheets

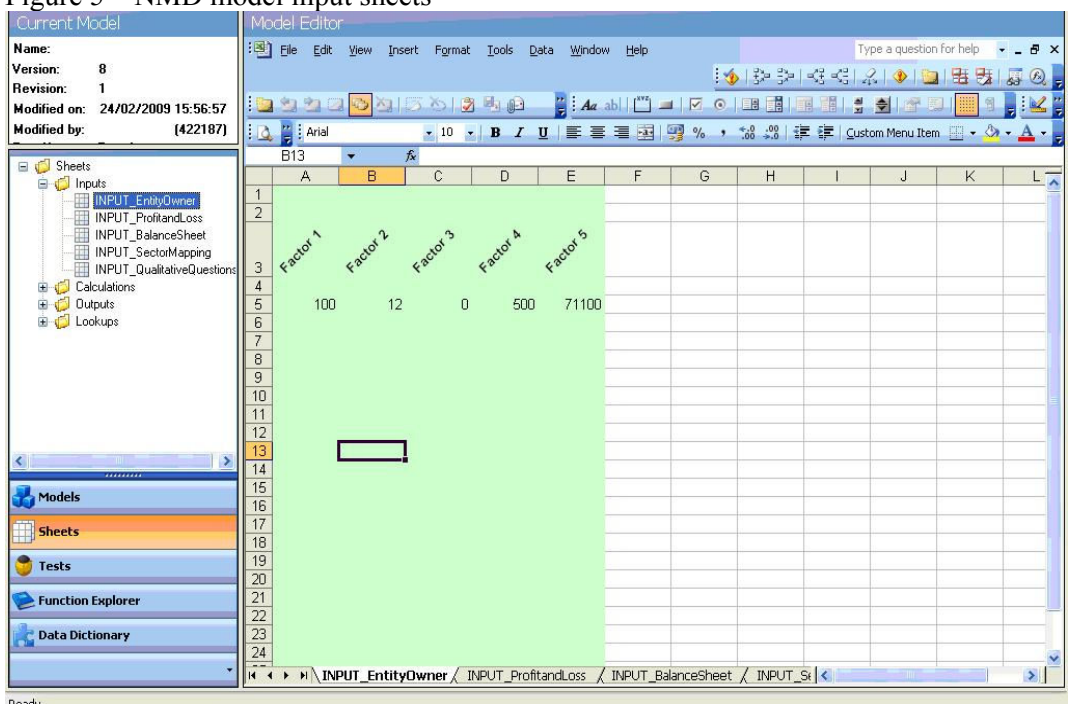

In addition the input sheets in the models follow the object structure observed in the Nexus database rather than the financial statement layout, again something with which the RMs are not immediately familiar.

The models can now be exported to all users, which is a major step forwards. However the planning around how the RMs and sanctioners would use this new toolset has been underestimated and needs further attention to ensure model build projects are a complete success. As a result of the feedback, the model developers have now instigated a phase in the model build projects where they will discuss the models with the users and discuss with them how the structures are presented. This alleviates any potential confusion around data structures and the look and feel of the models.

The NMD project was driven by the Risk Analytics Implementation team requirements and has satisfied many requirements in terms of building a tool which could interpret deals end to end, as well as having added further functionality around auditing. It is interesting to see that the people aspect [Grossman, 2002], whilst it may be frequently missed, is seen time and again across organisations. David Elton, in his article in the Financial Times [Elton, 2008], notes that IT is all about people and their reactions and feelings. He notes that "changing the way people work, not just the technology, has to be at the core" (of every IT project).

This is where the challenge now lies. The technology is now readily available, it is now necessary to ensure that all parties can see the benefits and adapt to looking at Excel in a slightly different way to their usual VB-adapted screens.

## 6. CONCLUSIONS

The NMD project has addressed most concerns raised by the various model users. The interface shown to the RMs is a minor issue and therefore an aspiration to look towards. Time spent up-front in model build projects to engage the RMs and explain the look of the NMD models will avoid time spent post model build convincing users of the benefits of using Excel in the NMD.

These benefits include ease of access to a system available to all to view the models, which is a major step forwards from previous model build systems which were viewed as 'black boxes'. There is now full transparency for the RMs through the model calculations using either simple Excel functions or in the more complex cases by using a wizard function i.e. for arrays.

In addition the system has been built to include a version control function which picks up all changes to Excel sheets, and differentiates between model inputs/outputs which will affect the model structure, and the internal calculations. This change control function provides full confidence in the auditing of all changes.